\documentstyle[emulateapj]{article}

\newcommand{\etal}{{et al. \,}}
\newcommand{\kms}{km~s$^{-1}$}

\def\gtsim{\ {\raise-0.5ex\hbox{$\buildrel>\over\sim$}}\ }
\def\ltsim{\ {\raise-0.5ex\hbox{$\buildrel<\over\sim$}}\ }

\lefthead{Lee  et al.}
\righthead{Dwarf Galaxy DDO 210}

\begin{document}

\title{STELLAR POPULATIONS AND THE LOCAL GROUP MEMBERSHIP \\
OF THE DWARF GALAXY DDO 210$^1$}

\author{Myung Gyoon Lee}
\affil{Department of Astronomy, Seoul National University, Seoul 151-742, 
Korea \\
Electronic mail: mglee@astrog.snu.ac.kr}

\author{Antonio Aparicio}
\affil{Instituto de Astrof\'\i sica de Canarias, La Laguna, Tenerife,
Canary Islands, Spain\\
Electronic mail: aaj@iac.es}

\author{Nikolay Tikonov}
\affil{Special Astrophysical Observatory, Satavropol, Russia\\
Electronic mail: ntik@nik.sao.ru}

\author{Yong-Ik Byun}
\affil{Department of Astronomy and Center for Space Astrophysics, 
Yonsei University, Seoul 120-749, Korea \\
Electronic mail: byun@darksky.yonsei.ac.kr}

\and
\author{Eunhyeuk Kim}
\affil{Department of Astronomy, Seoul National University, Seoul 151-742, 
Korea \\
Electronic mail: ekim@astro.snu.ac.kr}



\altaffiltext{1}{
Based on observations made with the 2.5 m Nordic Optical Telescope
operated on the island of La Palma by NOT S.A. in the 
Spanish Observatorio del Roque de Los Muchachos of the 
Instituto de Astrof\'\i sica de Canarias.}


\begin{abstract}

We present deep $BVI$ CCD photometry of the stars in the dwarf galaxy
DDO 210. The color-magnitude diagrams of DDO 210 show a well-defined red
giant branch (RGB) and a blue plume.  The tip of the RGB is found to be
at $I_{\rm TRGB} = 20.95 \pm 0.10$ mag.  From this the distance to DDO
210 is estimated to be $d = 950\pm 50 $ kpc.  The corresponding distance
of DDO 210 to the center of the Local Group is 870 kpc, showing that it
is a member of the Local Group.  The mean metallicity of the red giant
branch stars is estimated to be [Fe/H] = $-1.9\pm0.1$ dex.  Integrated
magnitudes of DDO 210 within the Holmberg radius ($r_H=110$ arcsec $ =505$ pc)
are derived to be $M_B=-10.6\pm0.1$ mag and $M_V=-10.9\pm0.1$ mag.  $B$
and $V$ surface brightness profiles of DDO 210 are approximately
consistent with an exponential law with scale lengths $r_s(B)$ = 161 pc
and $r_s(V)$ = 175 pc.  The brightest blue and red stars in DDO 210 (BSG
and RSG) are found to be among the faintest in the nearby galaxies with
young stellar populations: $<M_V(3)>_{BSG} = -3.47\pm 0.11$ mag and
$<M_V(3)>_{RSG} = -4.75 \pm 0.13$ mag.  An enhancement of the star
formation rate in the recent past (several hundred Myrs) is observed in
the central region of DDO 210.  The opposite trend is observed in the
outer region of the galaxy, suggesting a possible two-component
structure of the kind disk/halo found in spiral galaxies. The real
nature of this two-component structure must, however, be confirmed
with more detailed observations.

\centerline{[To appear in the Astronomical Journal in August 1999]}
\vskip 0.1in

\end{abstract}




%

\section{INTRODUCTION}

DDO 210 is a faint dwarf galaxy in Aquarius discovered by van den Bergh
 (1959). It was resolved in ground based photographic images
 so that
it was sometimes considered to be located nearby and to be a member of
the Local Group (\cite{fis75}, 1979; \cite{yah77}). It has been known
to be a dwarf irregular galaxy.
Lo \etal (1993)
 obtained an HI mass of $3 \times 10^6$ M$_\odot$ and a total virial mass
 of $1.4\times 10^7$ M$_\odot$ in DDO 210, calculated
 assuming a distance of 1.0 Mpc. Taylor \etal (1998) gave an upper limit
 for the CO emission of $I_{CO}<0.11$ K\,km\,s$^{-1}$.

The distance to DDO 210 has never been measured reliably
and is not yet well known. From the proximity of DDO
210 on the sky and in velocity to NGC 6822, Fisher \& Tully (1979)
assumed that both galaxies were at the same distance (0.7 Mpc). 
 On the other hand, from $BV$ CCD
photometry of bright stars in DDO 210, Greggio \etal (1993)
suggested
that this galaxy may be located at as far as 4 Mpc, placing it
beyond the edge of the Local Group. 
van den Bergh (1994) estimated the
distance to be 0.8 Mpc, from simple comparison
of the DDO 210 and the Leo I color-magnitude diagrams.

In this paper we present a study of stellar populations in DDO 210 and 
the Local Group membership of this galaxy based on deep ground
based $BVI$ CCD photometry.  
A brief progress report of this study was given in
Lee (1999), which is superseded by this paper. 
This paper is composed as follows.
Section 2 describes the observations and data reduction; 
Section 3 investigates the morphological structure of DDO 210. 
Section 4 presents the color-magnitude diagrams of  the galaxy; and
Section 5 estimates the distance to DDO 210 and concludes 
on the Local Group membership of DDO 210.
Section 6 presents the surface photometry of DDO 210,
Section 7 investigates the property of the brightest blue and red stars
in DDO 210,
and Section 8 discusses the star formation history (SFH).
Finally,  Section 9 summarizes the primary results.

\section{OBSERVATIONS AND DATA REDUCTION}

\subsection{Observations and Data Reduction}

$BVI$ CCD images of DDO 210 were obtained during three different observing runs
at the Palomar 1.5 m telescope, the University of Hawaii 2.2 m telescope,
and the NOT 2.5 m telescope of Roque de los Muchachos Observatory at La Palma.
Table 1 lists the journal of  observations of DDO 210.
A grey scale map of the $V$-band CCD image of DDO 210 obtained at UH 2.2 m
is displayed in Fig. 1  
which shows that there are many bright foreground stars
in the direction of this galaxy. 
For the analysis of the data, we have divided 
the field covered by our CCD images into three regions as shown in Fig. 1:
the C-region  covers the central main body of DDO 210, the I-region
roughly extends to the boundaries of the optical body of the
galaxy, the F-region is a control field and the F2-region  represents 
the field outside the I-region in the Roque de los Muchachos field. 

Instrumental magnitudes of the stars in the CCD images were derived
using DoPHOT (\cite{sch93}) and DAOPHOT(\cite{ste87}).  Palomar and
Roque de los Muchachos data were taken under photometric conditions 
and were calibrated using several standard stars (\cite{lan92}) observed
on the same runs. The calibration errors of each photometry are 0.04--0.05.
The magnitudes of a subset of bright, well measured stars in common between
 the Palomar and Roque de los Muchachos photometries were averaged to obtain a common
photometric scale to which all the magnitudes were transformed. The
Hawaii data were also transformed to this photometric scale.

The Roque de los Muchachos $VI$ photometry is the
deepest one so that we have  preferably
used it for the analysis of the stellar populations
in DDO 210. The Hawaii $BVI$ data were used for the analysis of the bright
stellar populations.
The total number of stars which were measured in $V$ and at least one other
color with ALLSTAR $\sigma$ values less than 0.25 is $\sim 1800$.
 Table 2 lists $BVI$ photometry of the measured
bright stars with $V<22$ mag inside the boundary of the I-region. 
$X$ and $Y$ coordinates increase to the east and
to the south, respectively.

\subsection{Comparison with Previous Photometry}

Previous $BV$  photometries of stars in DDO 210 were given by
Marconi \etal (1990), 
Greggio \etal (1993, G93 from now on) and Hopp \& Schulte-Ladbeck (1995,  HP 95 from now on). 
The photometry  by Marconi \etal (1990) was
based on a wrong calibration and was corrected in G93.
 We have compared our photometry with these previous ones. 
The results of the comparison are
   $\Delta V$ (this study -- G93)     = $+0.01\pm 0.09$,
   $\Delta (B-V)$ (this study -- G93) = $+0.05\pm 0.15$  
                          for 30 common stars with $V<22$ mag, and
   $\Delta V$ (this study -- HP95) = $-0.04\pm 0.10$,
   $\Delta (B-V)$ (this study -- HP95) = $+0.05\pm 0.11$  
                          for 35 common stars with $V<22$ mag.
These results show a reasonable agreement in $V$ and $(B-V)$ 
among the three photometry sets.

\section{MORPHOLOGICAL STRUCTURE}

 Fig. 1 shows that, morphologically, DDO 210 seems 
to have two components: the central component, represented by the C-region
(of size $2'.32 \times 0'.87$), and the extended component, represented
by the I-region (of size $4'.85 \times 2'.13$). 
From now on, by the I-region, we will refer only to the outer part
of this region, i.e.,  excluding the C-region. 
Both are elongated in the east-west direction, but have not
common centers, the central component lying more east-ward than the
extended one.  A color-map created by combining $BVI$ images
shows that the stars in the central component are mostly very blue, 
while the stars in the outer region are mostly
yellow to red. We have estimated the ellipticity and the position
angle of the extended component to be roughly $e=0.5$ and 
$PA=173^{\rm o}$, respectively.
Summarizing, DDO 210 is an elongated dwarf galaxy 
with a likely young stellar component close but out of its center.
This confirms the common tendency of dwarf galaxies to show off-center star
forming regions.

\section{COLOR-MAGNITUDE DIAGRAMS}

We display $V$--$(B-V)$ and $I$--$(V-I)$ diagrams of the measured stars in the
C-region, I-region and F-region in Figs. 2 and 3. 
Note that the area of the (C+I)-region in the field is the same 
as that of the F-region so that we can estimate the contamination due to
foreground stars by comparing the diagrams of each region.

Several distinguishable features of the stars in DDO 210 are seen 
in Figs. 2 and 3.
First, there is a blue plume of bright stars with $(B-V)<0.4$ in the
(C+I) region. Comparison of the (C+I)-region and the F-region in Fig. 2
 shows that these bright blue stars  are mostly  members of DDO 210.
The brightest end of the blue plume extends up to $V\approx 21.0$ mag and
$(B-V)\approx -0.2$. These stars are young stars and are mostly
located in the C-region. 

Secondly, there are several bright red stars with $20<V<21$ mag and
$1.0<(B-V)<1.5$ in the C and I-region in Fig. 2. They might be young,
red supergiants in DDO 210. However, the fact that a large number of
stars is found in the F-region with $V < 22$ mag (for every color);
the fact that they are more numerous in the I-region than in the C-region, 
as it
should be expected from the bigger size of field; and, more important,
the fact that the blue population (which must be expected as a
counterpart of the bright, red one) in the I-region is less numerous than in
the C-region, would drive to the conclusion that many of the bright, red
stars in field I are foreground.

Thirdly, Fig. 3 shows that there is a strong concentration of red stars
fainter than $I \approx 21$ mag and redder than $(V-I)\approx 0.8$ mag
in the C and I-regions. This feature corresponds to the red giant branch
(RGB) and asymptotic giant branch (AGB)
locii, populated by old and intermediate-age stars of DDO 210.
The tip of the RGB (TRGB) is seen at $(V-I) \approx 1.4$.

\section{DISTANCE AND METALLICITY}

We estimate the distance to DDO 210 using the $I$ magnitude of the TRGB, 
as described in Da Costa \& Armandroff (1990) and Lee \etal (1993). 
The $I$ magnitude of the TRGB is estimated
using the $I-(V-I)$ diagram in Fig. 3 and the luminosity function of red
stars in the range defined by $(V-I, I) =$ (0.5, 23.5), (1.2, 23.5) and (1.8, 19).
Fig. 4 shows the $I$-band luminosity function of the measured red stars 
in the (C+I)-region. We also plot the luminosity functions of stars
of the same color range
in the F-region and the F2-region in Fig. 4. This figure shows that while
some star of DDO 210 can be still populating the F2-region, the F-region
is very likely free of them. In any case, for the purposes of this
section, the contribution due to foreground stars is negligible
for the bright part of the luminosity function of the (C+I) region.
The luminosity function
shows a sudden increase at $I=20.95\pm0.10$ mag. This corresponds to
the TRGB seen in the color-magnitude diagram in Fig. 3. We will adopt
this value as the $I$ magnitude of the TRGB.

The mean color of the TRGB is estimated to be $(V-I)=1.44\pm 0.04$.
The bolometric magnitude of the TRGB is then calculated from
$M_{\rm bol}=-0.19{\rm [Fe/H]} - 3.81$.
Adopting a mean metallicity of [Fe/H] = $-1.86\pm 0.12$  dex as 
estimated below, we obtain a value for 
the bolometric magnitude of $M_{\rm bol}=-3.46$ mag.
From the equation BC$_I$ = 0.881 -- 0.243$(V-I)_{\rm TRGB}$ 
 the bolometric correction at $I$ for the TRGB is estimated 
to be BC$_I=0.54$ mag.
The intrinsic $I$ magnitude of the TRGB is then
$M_I= M_{\rm bol} - {\rm BC}_I = -4.00$ mag.

We adopt in this study the foreground reddening value
$E(B-V)=0.03$  mag for DDO 210 in Burstein \& Heiles (1982)
and Ratnatunga \& Bahcall (1985), corresponding to $A_I=0.06$ mag. This
value is also similar to that 
given by Schlegel \etal (1998), $E(B-V)=0.05$.
Finally the distance modulus of DDO 210 is obtained:
$(m-M)_0 = 24.89 \pm 0.11$ mag,
corresponding to a distance of $950\pm 50$ kpc .

The metallicity of an old stellar population can be estimated from
the color of the RGB. DDO 210 has a composite stellar population
containing also young stars, but an estimate of the metallicity using
that approach is still quite useful, since no previous determinations
exist. We have estimated the mean metallicity 
 using the $(V-I)$ color of the RGB stars 0.5 mag fainter
than the TRGB, $(V-I)_{-3.5}$.
This color is measured from the median value of the colors of red giant
branch stars with $I=21.45\pm0.05$ mag in the I-region, resulting to be
$(V-I)_{-3.5} = 1.34 \pm 0.04$. From this value
we estimate the mean metallicity to be [Fe/H] $= -1.86\pm 0.12$ dex.
This value is similar to those of dwarf irregular galaxies 
Sextans A ([Fe/H] $= -1.9 \pm 0.1$ dex)
and UKS 2323--326 ([Fe/H] $= -1.9 \pm 0.1$ dex),
and is close to the lowest end in the metallicity of stellar
populations in dwarf irregular galaxies (\cite{mat98}, \cite{lee99b}).

In Fig. 5 we overlayed the loci of the red giant branches of
Galactic globular clusters, of M15, M2, and NGC 1851, shifted according to
the distance and reddening of DDO 210.  
The metallicities of M15, M2 and NGC 1851 are 
[Fe/H] = --2.17, --1.58 and --1.29 dex, respectively.
Fig. 5 shows that the bright part of the RGB of DDO 210 is located
well between those of M15 and M2. The broadening of the faint part of the RGB
is mostly due to the photometric errors.


Greggio \etal (1993) estimated the distance to DDO 210 to be 4 Mpc, which
is four times larger than our result. They derived this
value  from the comparison of their $BV$ photometry of DDO 210
with the synthetic color-magnitude diagrams created using the Padova
models of stellar evolution. They  stated in their
paper that ``the true distance
modulus of DDO 210 is $(m-M)_0 \approx 28$, although the uncertainty is 
obviously large''. 
If DDO 210 were at the distance of 4 Mpc, 
the absolute magnitude of the bright tip of the RGB  would be
 $M_I \approx -7$ mag. This is unrealistic, because 
the $I$-band absolute magnitude of the TRGB is known to
be almost constant at $M_I \approx -4.0$ mag for various types of resolved
galaxies with old red giant populations  
(\cite{lee93}).

 With the distance estimate obtained as above, 
we can tell whether DDO 210 is a member 
of the Local Group or not.
Assuming that the center of the Local Group is,  from our Galaxy, 
at two thirds of the connection line between our Galaxy and M31,
we derive the distance to DDO 210 from the center of the Local Group to be 870 kpc. In  addition,
the systemic velocity of DDO 210 is $v = -137\pm5$ \kms (RC3), 
and the velocity with respect
to the center of the Local Group is derived to be --15 \kms. 
From these values we conclude that DDO 210 is  definitely
 a member of the Local Group.  
DDO 210 does not belong to either of the Milky Way subgroup or
the M31 subgroup.

\section{SURFACE PHOTOMETRY}

The low surface brightness of DDO 210 and the presence of several
bright foreground stars in the field, makes it difficult to derive reliably
the surface photometry of the galaxy. We have proceeded for the surface
photometry as follows.
First, we removed several bright stars which are considered obviously to be
foreground stars both from the DDO 210 and the F-region images 
using IMEDIT in IRAF. 
The mean intensity of the F-region was used for background sky estimation.
Then we performed aperture photometry of DDO 210 using elliptical annular apertures
 with the ellipticity
and position angle determined in Section 3. 
The results are listed in Table 3 and 
are displayed in Fig. 6.
In Table 3 $r_{eff}$ represents  a mean radius of the major axis of an ellipse, and $r_{out}$ presents the outer radius of an elliptical aperture.
Surface brightnesses are given in terms of mag arcsec$^{-2}$.

Fig. 6  shows that the surface brightness profiles of DDO 210 
follow approximately the exponential disk law. 
The scale lengths of DDO 210 are determined from these profiles
 to be $r_s (B)= 35 $ arcsec = 161 pc and  $r_s (V)= 38 $ arcsec = 175 pc.
The steep increase of the brightness
in the center is due to a bright star located close to the center of the galaxy.
The magnitude and color of this star (ID 1407) are $V=20.87\pm0.10$
and $(B-V) = 1.14\pm0.10$.  With this color, it could be a main sequence star
in the halo of the Milky Way, 5 kpc from the Sun or 7.5 kpc from the
galactic center. It is also possible, however, that it is a yellow
supergiant in DDO 210.
 The color profile of DDO 210 in Fig. 6 shows that the colors are
almost constant around $(B-V) \approx 0.25$ within $r \approx 50$
arcsec (except for the central region), 
and get redder beyond $r \approx 50$ arcsec. This shows that
blue stellar populations are located mostly within $r \approx 50$ arcsec.

The Holmberg radius corresponding to $\mu_B=26.5$ mag arcsec$^{-2}$ is
approximately $r_H=110$ arcsec = 505 pc. The integrated magnitudes of DDO
210 within this radius are:
$B=14.37\pm0.03$ mag, $V=14.06\pm0.03$ mag  and $I=13.03\pm0.04$ mag.
de Vaucouleurs \etal (1991, RC3) lists, as total magnitudes of DDO 210,
$B=14.00 \pm 0.52$ mag and $V=13.88 \pm 0.52$ mag.
These magnitudes are similar to ours, but the errors in our photometry
are much smaller.
The corresponding absolute magnitudes are $M_B = -10.6\pm0.1$
mag, $M_V = -10.9\pm 0.1$ mag, and $M_I = -11.9\pm0.1$ mag.
This result shows that DDO 210 is the faintest among the known
dwarf irregular galaxies in the Local Group (\cite{mat98}).

\section{THE BRIGHTEST BLUE AND RED STARS IN DDO 210}

Little is known about the nature of the brightest stars in DDO 210. 
In the study of M supergiants in Local Group irregular galaxies 
Elias \& Frogel (1985) stated ``Certainly DDO 210 does not have a population
of red supergiants comparable to that in Sextans A, which implies that it either
is more distant or has no red supergiants.''.
Since it turns out that DDO 210 is closer than Sextans A 
(this study, \cite{mat98}), 
the only possibility is that there may be no red supergiants in DDO 210.
Our photometry allow us to investigate the nature of the brightest stars 
in DDO 210.

We have determined the mean colors and magnitudes of the brightest
stars.  In the $V-(B-V)$ diagram of the (C+I) region of DDO 210 shown in
Fig. 2 it is obvious which stars are the three brightest blue stars in
DDO 210 (IDs: 1656, 1503, and 1517), but it is not obvious which stars
are the three brightest red stars because of contamination due to some
foreground stars.  We chose, as the three brightest red stars, 
three among the ones with
$20<V<21$ and $1.0<(B-V)<1.5$ (IDs: 1468, 1071, and 926).
The mean magnitudes and colors of these three brightest blue and red
stars (called as BSG and RSG hereafter) in DDO 210 are derived to be,
respectively, $<V(3)>_{BSG} = 21.51\pm0.11$ mag and $<(B-V)(3)>_{BSG} =
-0.15\pm0.03$, and $<V(3)>_{RSG} = 20.23\pm0.13$ mag and
$<(B-V)(3)>_{RSG} = 1.13\pm 0.12$.  The corresponding absolute
magnitudes and colors are $<M_V(3)>_{BSG} = -3.47$ mag, $<(B-V)(3)>_{0,
BSG} = -0.18$, $<M_V(3)>_{RSG} = -4.75$ mag and $<(B-V)(3)>_{0, RSG}
= 1.10$, respectively.
If some of the chosen brightest red stars are foreground stars, 
the mean magnitudes of the
brightest red stars in DDO 210 will be fainter.

We have compared the magnitudes and colors of the brightest stars in DDO 210
with those of other galaxies with young stellar populations in Fig. 7. 
The data for other galaxies are from Lyo \& Lee (1997).
Fig. 7 shows that both the brightest blue and red stars in DDO 210 are 
the faintest among the sample galaxies. 
This result is consistent with the fact that DDO 210 is the faintest
among the known dwarf irregular galaxies in the Local Group.
Fig. 7 shows also that  the brightest red stars
are the least red among the sample galaxies. This can be interpreted
in the sense that
the brightest red stars in DDO 210 may be more metal-poor  than
 those in brighter galaxies. 

\section{THE STAR FORMATION HISTORY OF DDO210}


From a qualitative point of view, the presence of a conspicuous
RGB-AGB structure implies star formation at intermediate to old ages in
DDO 210. Moreover, the age of the youngest population in DDO 210 can be
estimated from comparison of the CMD of Fig. 2 with theoretical
isochrones. Fig. 8 displays the $V-(B-V)$ CMD of DDO 210 together with
three isochrones for metallicities $Z=0.0004$ and ages 30, 100
and 300 Myrs from the Padova library (see Bertelli \etal 1994 and
references therein). It shows that star formation as recent as some
30 Myrs ago has taken place in the galaxy.]

The star formation history (SFH) can be derived quantitatively in detail
from deep CMDs as shown by Gallart \etal (1999). The CMD of DDO 210 is
not deep enough for such a derivation, but it can still be used to
sketch the SFH. For this purpose, stars with $-4<M_{I,0}<-2.5$ have been
counted in the CMD of fields C-region, I-region
and F2-region and divided into two groups: bluer and redder than
$(V-I)_0=0.5$. After normalization to the same area, the counts of the
F2-region have been subtracted from those of the C and I-regions to
correct of foreground contamination. The relatively strong foreground
contamination of the DDO 210 field prevents us from using the stars over
the TRGB for the calculation of the SFH. On the other hand, the maximum
magnitude used ($M_{I}=-2.5$ mag) and the fact that we pursue only a rough
estimate of the SFR, allow us to neglect crowding effects. 

Following a method similar to that described in Aparicio \etal (1997b)
and Aparicio \etal (1999), 
a synthetic CMD has been calculated with 100,000 stars brighter
than $M_I=-1.5$ mag, a constant SFR and random metallicities in the interval
$0.0001<Z<0.0003$, which corresponds to the interval of [Fe/H] found
from the RGB of DDO 210. The synthetic stars have been divided into two
age intervals: younger and older than 1 Gyr, 
and the same boxes as for the observational CMDs
have been defined in each of the resulting synthetic CMDs. The resulting
SFRs for the 15 to 1 Gyrs ago and the 1 to 0 Gyrs ago are given in Table 4
 and shown in Fig. 9 for
the central (C) and the outer (I) regions of DDO 210. These values
are also given after surface normalization (right-hand ordinate axis of
Fig. 9). The areas used for those normalizations have been 2.18 arcsec$^2$
or $1.67\times 10^5$ pc$^2$ for the C-region and 8.11 arcsec$^2$ or
$6.21\times 10^5$ pc$^2$ for the I-region. 

DDO 210 shows a SFH behavior in its central region similar to what has
been found in other dIrr galaxies,
like NGC 6822 (\cite{gal96}), Pegasus (\cite{apa97a}), Antlia
(\cite{apa97c}) and DDO 187 (\cite{apa99}). In overall, it consists in
an enhancement of the SFR for the recent past of the history of the galaxy
with respect to the value averaged for its entire life. This can be
interpreted as the result of a fluctuating
star formation rate (see \cite{apa99}). It is also interesting to note
that this recent strength of the SFR does not affect  the entire
galaxy but the central part only. For the outer I-region, the SFR
decreased for recent epochs. This behaviour is common to other dIrr: WLM
(\cite{min96}), Antlia (\cite{apa97c}), Phoenix (\cite{mar99}), DDO 187
(\cite{apa99}) and may suggest a two-components structure of the kind
disk-halo of bigger spiral galaxies. However, more detailed data are
needed to firmly state such conclusion.

\section{SUMMARY AND CONCLUSION}

We have presented a study of the stellar populations and the Local Group
membership of the dwarf galaxy DDO 210
 based on deep $BVI$ CCD photometry.
The primary results obtained in this study are summarized as follows.

(1) $BVI$ color-magnitude diagrams of $\sim$1800 stars in the $7'.5\times 7'.5$
   area of DDO 210 have been presented.
 These color-magnitude diagrams  exhibit a well-defined RGB and a blue plume.

(2) The TRGB is found to be at $I=20.95\pm 0.10$  mag 
and $(V-I)=1.44\pm 0.04$ mag.
From this, a distance modulus of $(m-M)_0=24.89\pm 0.11$ mag is
derived, corresponding to a distance of $950\pm 50$ kpc. The distance to
the barycenter of the Local Group is derived to be  870 kpc.
From this result and the systemic velocity of DDO 210 
%
we conclude that DDO 210 is definitely a member of the Local Group.

(3) The mean color of the RGB at $M_I=-3.5$ mag is $(V-I)=1.34\pm 0.04$
mag.  From this value we obtain a mean metallicity of the RGB: [Fe/H]
= $-1.86\pm0.12$ dex.

(4) $B$ and $V$ surface brightness profiles of DDO 210 follow
 roughly the exponential disk law with scale lengths of $r_s (B)= 35
 $ arcsec = 161 pc and  $r_s (V)= 38 $ arcsec = 175 pc. The Holmberg radius
 corresponding to $\mu_B=26.5$ mag arcsec$^{-2}$ is $r_H = 110$ arcsec = 505
 pc. The integrated magnitudes up to this radius are 
$M_B = -10.6$ mag, $M_V = -10.9$ mag, and $M_I = -11.9$ mag.

(5) The magnitudes of the brightest blue and red stars in DDO 210
are derived: $<M_V(3)>_{BSG} = -3.47 \pm 0.11$ mag and 
$<M_V(3)>_{RSG} = -4.75 \pm 0.13 $ mag. 
The brightest blue and red stars in DDO 210 are found to be 
among the faintest in nearby galaxies with young stellar populations.

(6) The SFRs of the central and extended regions of DDO 210 have
been calculated. The central region shows an
enhancement of the star formation activity in recent epochs, in
agreement to what has been observed in other dwarf irregular
galaxies. In the extended region of DDO 210, the SFR has been lower in
the recent past than the value averaged for the entire life of the
galaxy, indicating a possible two-components structure for the
galaxy that must, however, be confirmed with more detailed data.

\acknowledgments

This research is supported by
the Ministry of Education, Basic Science Research Institute grant 
No.BSRI-98-5411 (to M.G.L.); by the Instituto de Astrof\'\i sica de
Canarias, grant PB3/94 and the DGES of the Kingdom of Spain, grant
PB97-1438-C02-01 (to A.A.);
by Creative Research Initiatives Program of the Korean Ministry
of Science and Technology and also by Yonsei University Research Grant (to Y.I.B.).


%

%
%

\clearpage


\begin{figure}[1] 
\plotone{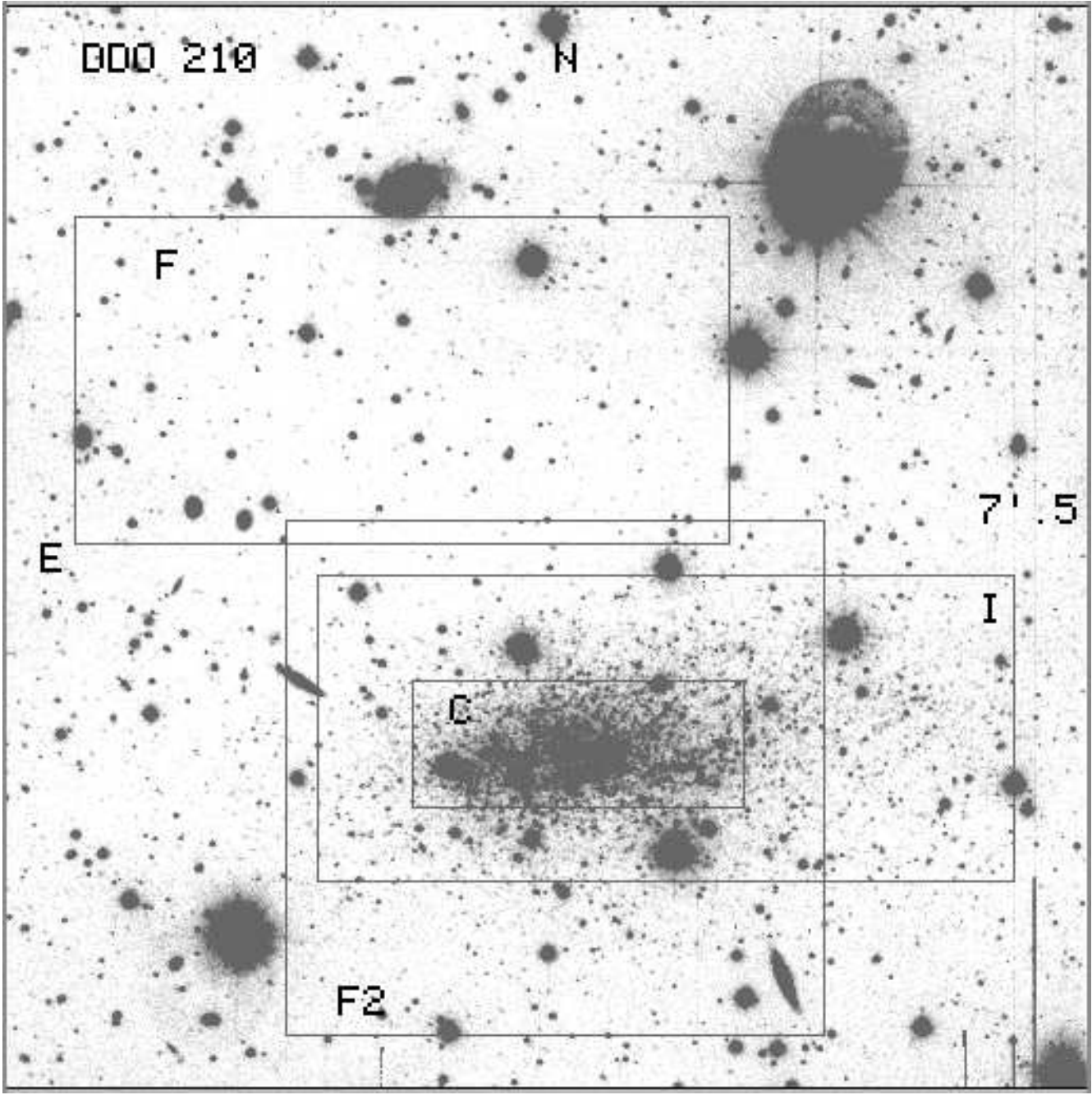}
\figcaption{
A grey-scale map of the $V$-band CCD image of DDO 210. 
North is at the top and east is to the left.
The size of the field is $7'.5 \times 7'.5$. 
Rectangular regions labeled as C, I, and F represent
 respectively,
the central region, the intermediate region and the control field region.
The square labeled F2 in the lower middle area represents the field covered
by the Roque de los Muchachos observations.
}
\end{figure}

\begin{figure}[2] 
\plotone{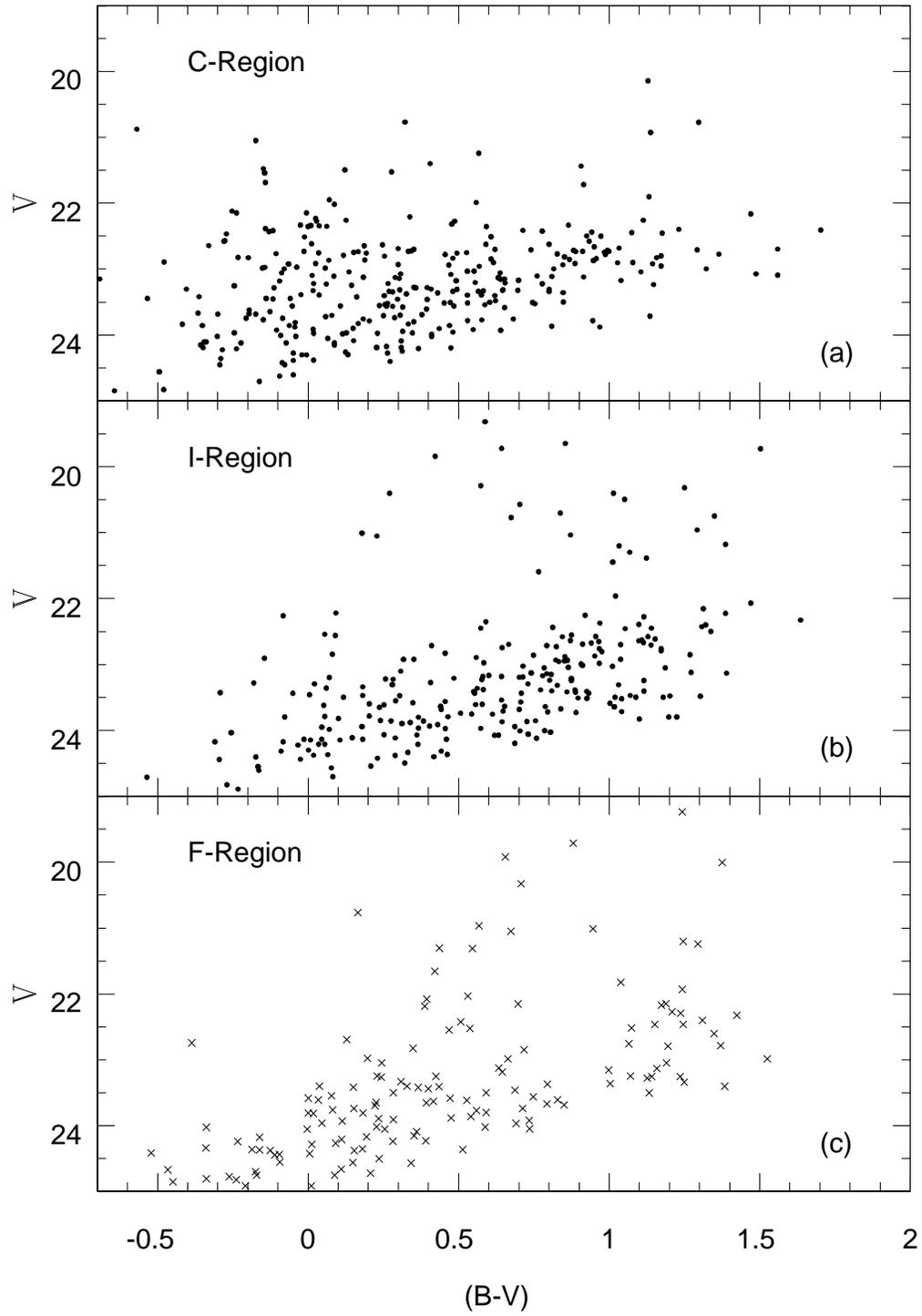}
\figcaption{$V$--$(B-V)$ diagram of the measured stars in the C-region  (a),
the I-region (b) and the F-region (c) of DDO 210. }
\end{figure}

\begin{figure}[3] 
\plotone{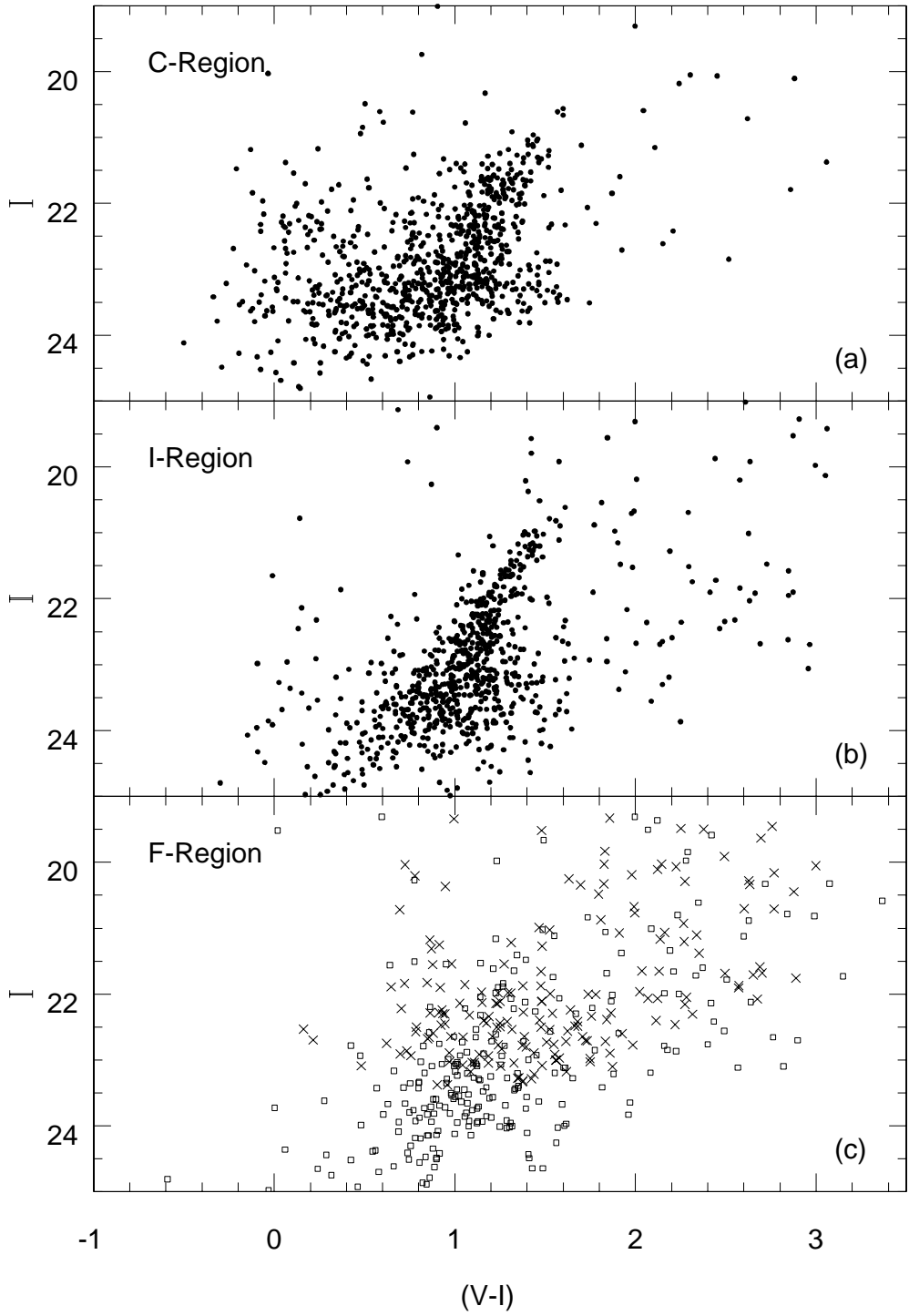}
\figcaption{
(a) $I$--$(V-I)$ diagram of the measured stars in the C-region (a), the I-region (b)
and the F-region (c) of DDO 210.
Open squares in (c) represent the stars in the F2-region.
}
\end{figure}

\begin{figure}[4] 
\plotone{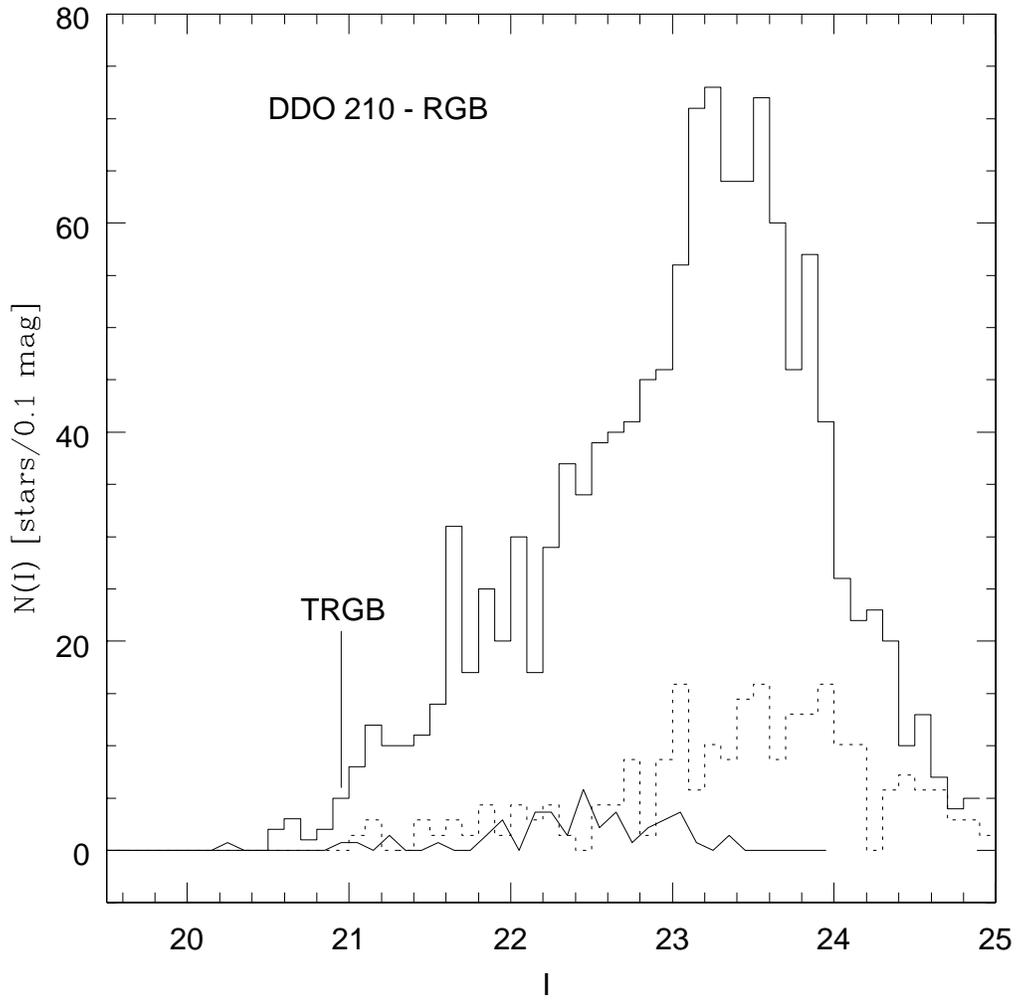}
\figcaption{
$I$-band luminosity function of the red giant branch stars in the C region
plus I-region. The tip of the red giant branch is labeled as TRGB.
The solid line and dotted lines represent the luminosity functions of
the F-region and F2-region, respectively.}
\end{figure}

\begin{figure}[5] 
\plotone{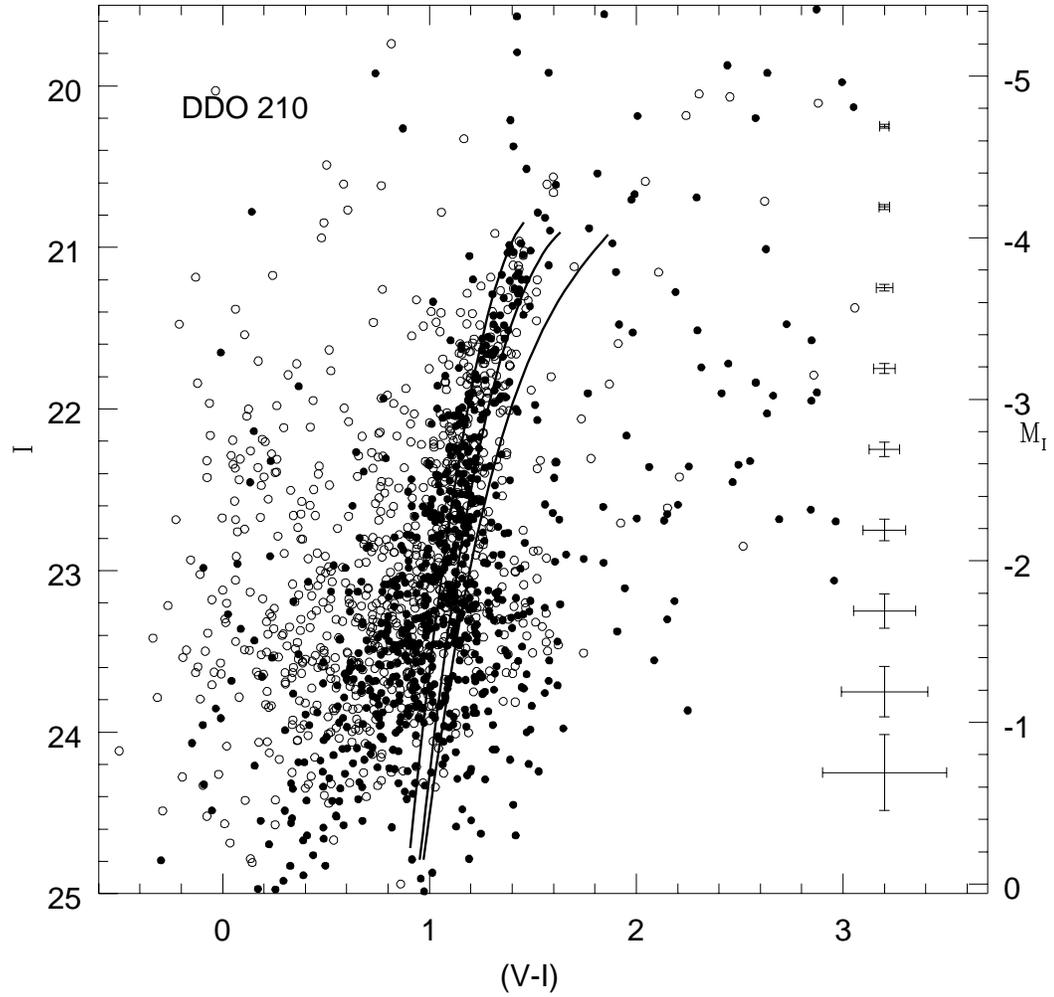}
\figcaption{
$I$--$(V-I)$ diagram of the measured stars in the I-region (filled circles) and
C-region (open circles) of DDO 210 in comparison with
the red giant branches of Galactic globular clusters. 
The solid curved lines show,
from left to right, the loci of the giant branches of M15, M2, and NGC 1851, 
the metallicities of which are [Fe/H] = --2.17, --1.58 and --1.29 dex, respectively.
The mean errors for the magnitudes and colors are illustrated 
by the error bars at the right.
}
\end{figure}

\begin{figure}[6] 
\plotone{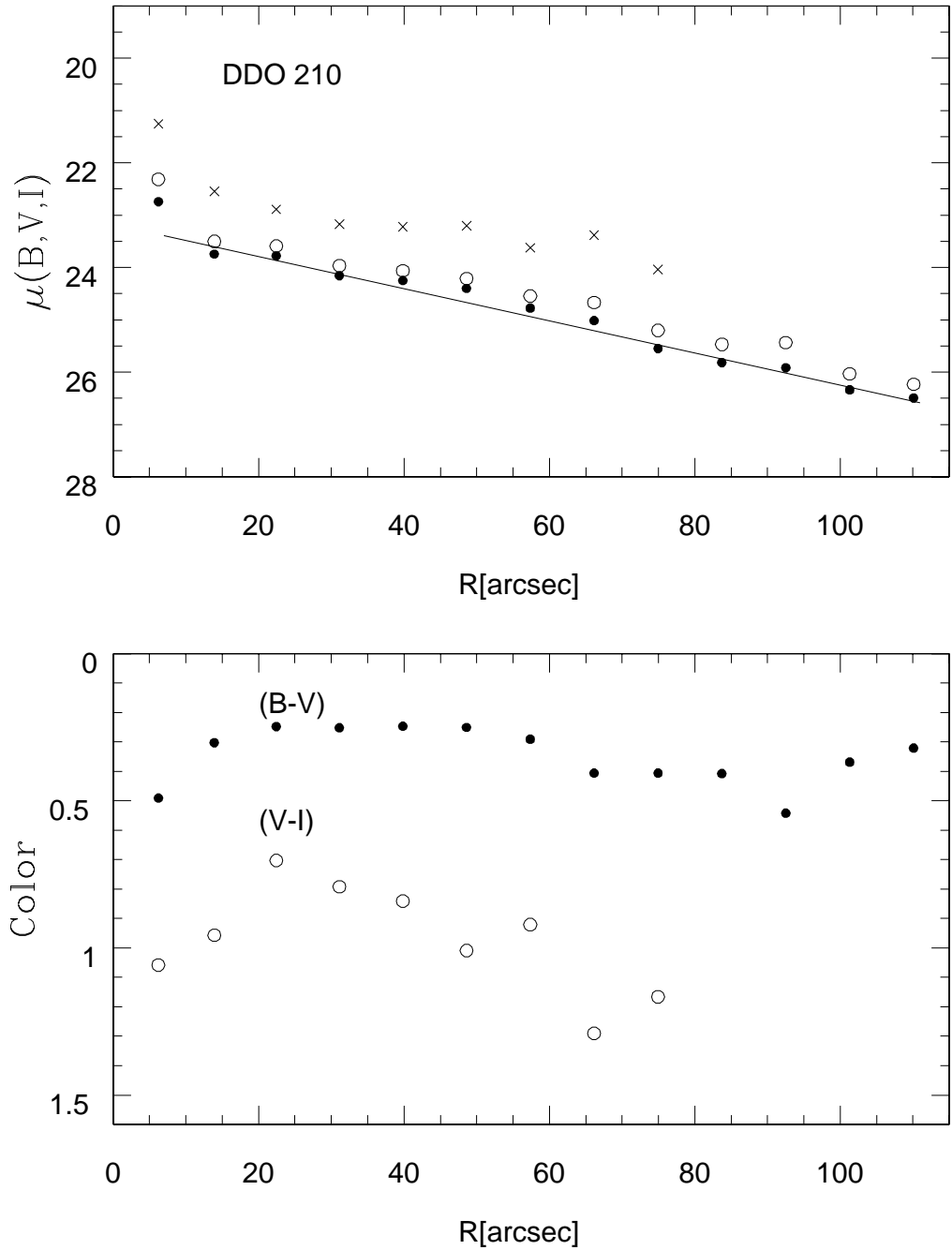}
\figcaption{ Surface photometry of DDO 210.
(a) Surface brightness profiles vs. effective radius along the major axis.
$B,V$ and $I$ magnitudes are represented by the filled circles, open circles 
and crosses. The solid line represents a linear fit to the $B$ data.
(b) Differential colors vs effective radius along the major axis.
}
\end{figure}

\begin{figure}[7] 
\plotone{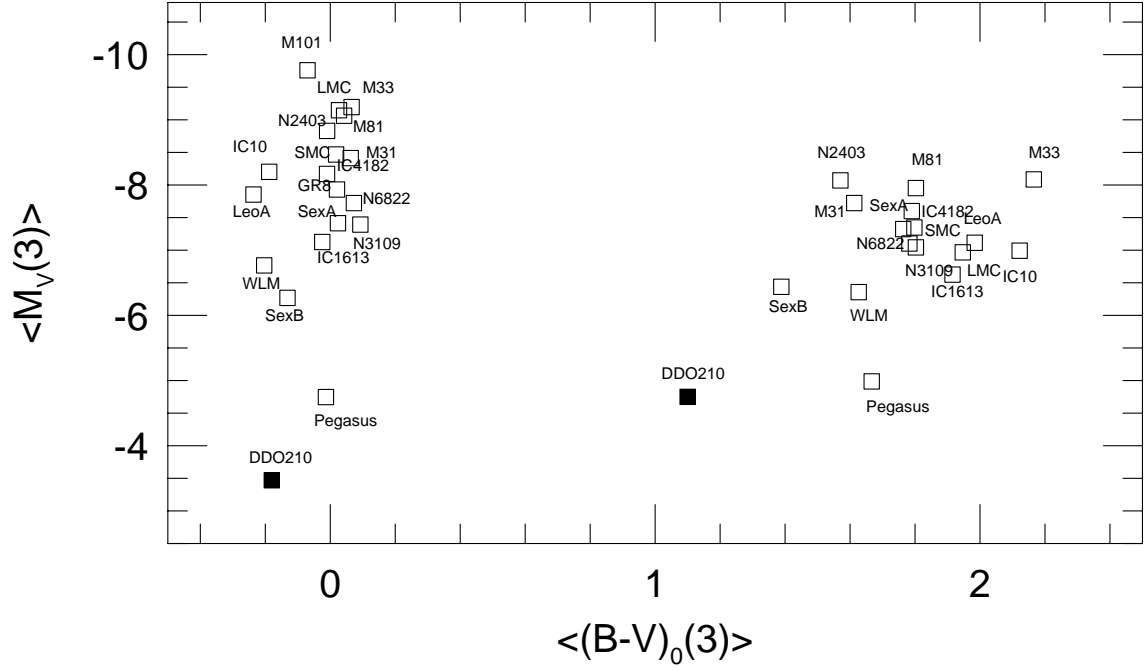}
\figcaption{ 
$V$--$(B-V)$ diagram of the brightest blue and red
supergiants in DDO 210 (filled squares) in comparison with other galaxies
(open squares). 
The magnitudes and colors are average values
of the three brightest blue and red supergiants in each sample galaxy.
}
\end{figure}

\begin{figure}[8] 
\plotone{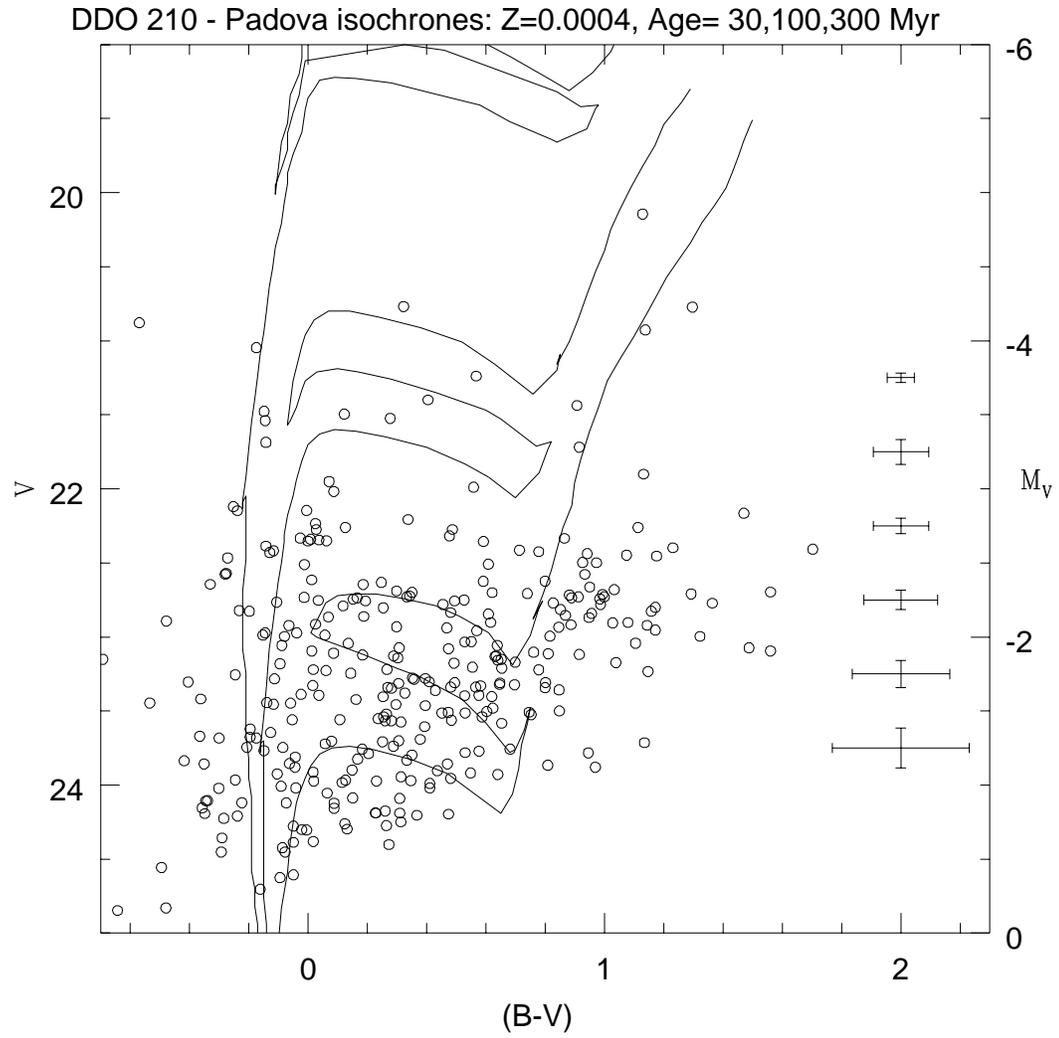}
\figcaption{ 
$V$--$(B-V)$ diagram of the measured stars in the C-region (open circles) of DDO 210 in comparison with theoretical isochrones.
The solid lines represent the Padova isochrones for the metallicity of 
$Z=0.0004$                 
and ages of 30, 100, and 300 Myrs.
The mean errors for the magnitudes and colors are
illustrated by the error bars at the right.
}
\end{figure}

\begin{figure}[9]
\plotone{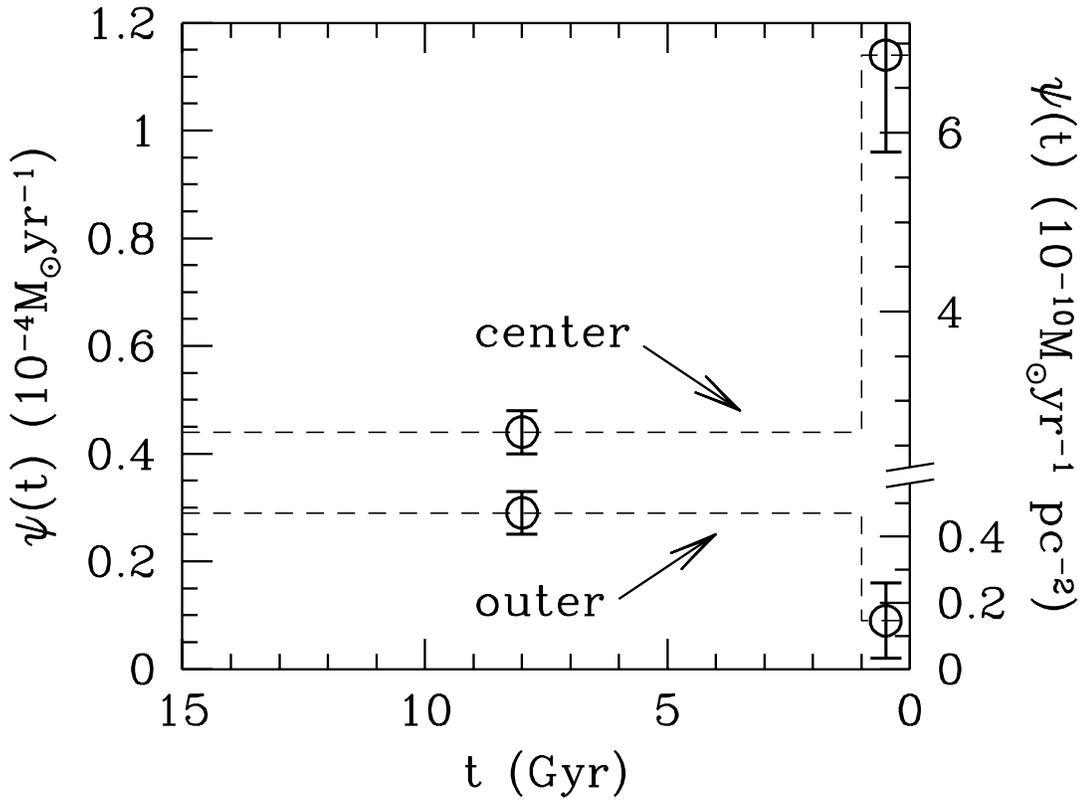}
\figcaption{
The star formation history of DDO 210. The bottom part of the
right-hand ordinate axis scale refers to the surface normalized
$\psi(t)$ of the external region of the galaxy. The upper part of the
same axis refers to the central region. The values of $\psi(t)$ for the
two age intervals plotted (15--1 and 1--0 Gyr ago) must be considered as
the averages for those intervals. The error bars correspond to
poissonian statistics on the star counts. The dashed lines show the
intervals of time over which the averages have been done, but are not
intended to give accurate information about the detailed shape of the
star formation history.
}
\end{figure}

\clearpage


%







\begin{deluxetable}{cccccl}
\tablecaption{JOURNAL OF OBSERVATIONS FOR DDO 210 \label{tbl-1}}
\tablehead{
\colhead{Filters} & \colhead{T$_{\rm exp}$}   & \colhead{Air mass}
& \colhead{FWHM} & \colhead{U.T.(Start)} & \colhead{Telescope}}
\startdata
$B$   &  1200 s       &  1.56 & $1''.2$ & 1992 Aug 23  07:52 & Palomar 1.5 m \nl
$V$   &  900 s & 1.65 & $1''.0$ & 1992 Aug 23 08:12  &    \nl
$I$   &  600 s & 1.45 & $0''.9$ & 1992 Aug 23 06:25 & \nl
$B$   &  1000 s       &  1.33 & $1''.2$ & 1994 Oct 9 07:32 & UH 2.2 m \nl
$V$   &  $3 \times 900$ s & 1.20 & $1''.0$ & 1994 Oct 9 06:05  &    \nl
$I$   &  $3 \times 600$ s & 1.25 & $0''.9$ & 1994 Oct 9 06:50 & \nl
$V$   &  $3 \times 900$ s & 1.21 & $0''.8$ & 1997 Jul 27 00:36 & NOT 2.5 m  \nl
$I$   &  $4 \times 600$ s & 1.26 & $0''.9$ & 1997 Jul 27 01:10 & \nl
\enddata
 
\end{deluxetable}





\begin{table*}
\begin{center}
\centerline{T{\small ABLE} 2.}
\centerline{ PHOTOMETRY OF THE BRIGHT STARS WITH $V<22$ MAG IN DDO 210}
\bigskip
\begin{tabular}{cccccc || cccccc}
\tableline\tableline
ID & X(pixel) & Y(pixel) & $V$ & $(B-V)$ & $(V-I)$ &
ID & X(pixel) & Y(pixel) & $V$ & $(B-V)$ & $(V-I)$ \\
\tableline
  1782&  526.3&  789.4&  16.85&   0.61&   0.83&  1407&  462.0&  702.5&  20.93&   1.14&   1.35\nl 
  1029&  405.3&  641.8&  17.09&   0.88&   1.06&  1829&  118.1&  814.4&  20.96&   1.29&   2.24\nl 
  1755&  358.8&  778.7&  17.19&   0.50&   0.87&  1362&  317.2&  694.8&  21.01&   0.18&   0.22\nl 
  1149&  300.1&  661.4&  17.48&   0.64&   0.84&  1656&  487.9&  745.8&  21.05&  -0.17&   0.04\nl 
  1571&  538.4&  726.6&  17.88&   0.55&   0.38&  1710&  468.5&  761.3&  21.05&   0.23&   0.63\nl 
  1423&  461.5&  704.5&  18.12&   0.86&   1.02&  1189&  668.9&  667.5&  21.20&   1.03&   0.85\nl 
  1763&  598.6&  783.2&  18.35&   0.90&   1.35&  1548&  441.5&  722.1&  21.24&   0.57&   0.61\nl 
  1202&  668.3&  669.4&  18.45&   0.90&   1.13&   799&  176.6&  558.5&  21.30&   1.07&   1.67\nl 
   855&  611.0&  590.1&  19.32&   0.59&   0.73&   910&  675.6&  620.1&  21.39&   1.12&   1.71\nl 
   918&  668.2&  622.5&  19.72&   0.64&   0.91&  1477&  461.5&  709.6&  21.40&   0.41&   0.72\nl 
  1243&  323.0&  675.7&  19.73&   1.50&   2.23&  1506&  470.3&  714.7&  21.44&   0.91&   1.15\nl 
  1780&  287.7&  788.0&  19.84&   0.42&   0.77&   820&  291.7&  571.0&  21.45&   1.01&   1.48\nl 
  1468&  490.5&  708.8&  20.14&   1.13&   1.74&  1503&  464.4&  713.9&  21.48&  -0.15&   0.19\nl 
  1805&  303.0&  801.4&  20.29&   0.57&   0.94&  1632&  515.3&  739.2&  21.50&   0.12&   0.62\nl 
  1071&  685.8&  647.1&  20.32&   1.25&   2.14&  1517&  523.8&  717.2&  21.54&  -0.14&   0.34\nl 
   926&  425.1&  625.0&  20.40&   1.01&   1.64&  1713&  321.5&  761.6&  21.59&   0.77&   1.38\nl 
  1444&  231.0&  706.2&  20.40&   0.27&   0.73&  1524&  560.7&  718.2&  21.69&  -0.14&  -0.10\nl 
  1830&  255.9&  815.4&  20.57&   0.70&   1.22&  1339&  563.5&  690.8&  21.72&   0.92&   0.49\nl 
  1821&  252.7&  811.5&  20.70&   0.84&   1.27&  1337&  410.0&  690.5&  21.90&   1.13&   1.62\nl 
  1206&  171.2&  669.8&  20.75&   1.35&   2.24&  1353&  452.4&  693.3&  21.95&   0.07&   1.28\nl 
  1600&  366.1&  733.0&  20.77&   0.32&   0.89&  1831&  424.8&  815.9&  21.96&   1.02&   1.51\nl 
  1551&  368.7&  722.3&  20.77&   1.30&   2.36&  1344&  420.6&  691.7&  21.99&   0.56&   0.61\nl  
\tableline

\end{tabular}
\end{center}
\end{table*}
\begin{deluxetable}{rccc rccc}
\tablenum{3}
\tablecaption{$BVI$ SURFACE PHOTOMETRY OF DDO 210
 \label{tbl-3}}
\tablewidth{0pt}
 \tablehead{ 
\colhead{$R_{\rm eff}$ [$''$]} & \colhead{$\mu_B$} & \colhead{$\mu_V$} & \colhead{$\mu_I$} &
\colhead{$R_{\rm out}$ [$''$]} & \colhead{$B$} & \colhead{$V$} & \colhead{$I$}  }
\startdata 
  $6.2$ & $22.74\pm0.43$ & $22.31\pm0.53$ & $21.26\pm0.68$ &   $8.8$ & $17.54$ & $17.08$ & $16.02$ \nl
 $13.9$ & $23.74\pm0.19$ & $23.50\pm0.23$ & $22.55\pm0.44$ &  $17.6$ & $16.69$ & $16.34$ & $15.32$ \nl
 $22.4$ & $23.78\pm0.31$ & $23.59\pm0.25$ & $22.89\pm0.23$ &  $26.4$ & $16.01$ & $15.73$ & $14.83$ \nl
 $31.1$ & $24.16\pm0.26$ & $23.96\pm0.22$ & $23.17\pm0.17$ &  $35.2$ & $15.59$ & $15.34$ & $14.47$ \nl
 $39.8$ & $24.25\pm0.27$ & $24.06\pm0.24$ & $23.22\pm0.18$ &  $44.0$ & $15.23$ & $15.00$ & $14.14$ \nl
 $48.6$ & $24.40\pm0.34$ & $24.21\pm0.31$ & $23.20\pm0.37$ &  $52.8$ & $14.95$ & $14.73$ & $13.83$ \nl
 $57.4$ & $24.77\pm0.33$ & $24.55\pm0.29$ & $23.62\pm0.32$ &  $61.6$ & $14.75$ & $14.53$ & $13.63$ \nl
 $66.1$ & $25.01\pm0.16$ & $24.67\pm0.19$ & $23.38\pm0.41$ &  $70.4$ & $14.61$ & $14.37$ & $13.40$ \nl
 $74.9$ & $25.55\pm0.24$ & $25.20\pm0.28$ & $24.04\pm0.58$ &  $79.2$ & $14.52$ & $14.27$ & $13.29$ \nl
 $83.7$ & $25.82\pm0.28$ & $25.47\pm0.38$ & $24.11\pm0.53$ &  $88.0$ & $14.44$ & $14.19$ & $13.20$ \nl
 $92.5$ & $25.92\pm0.33$ & $25.43\pm0.40$ & $23.77\pm0.60$ &  $96.8$ & $14.38$ & $14.11$ & $13.09$ \nl
$101.3$ & $26.34\pm0.40$ & $26.03\pm0.53$ & $24.82\pm1.03$ & $105.6$ & $14.33$ & $14.06$ & $13.03$ \nl
$110.1$ & $26.49\pm0.38$ & $26.23\pm0.60$ & $24.29\pm0.60$ & $114.4$ & $14.29$ & $14.02$ & $12.99$ \nl
$118.9$ & $26.27\pm0.49$ & $25.81\pm0.48$ & $24.26\pm0.51$ & $123.2$ & $14.23$ & $13.97$ & $12.92$ \nl
$127.7$ & $26.46\pm0.59$ & $26.03\pm0.48$ & $24.52\pm0.62$ & $132.0$ & $14.19$ & $13.92$ & $12.86$ \nl
\enddata
\end{deluxetable}

%


\clearpage
%

\begin{deluxetable}{lcc}
\tablenum{4}
\tablecaption{STAR FORMATION RATES \label{tbl-4}}
\tablewidth{0pt}
 \tablehead{ 
\colhead{} & \colhead{DDO 210 center} & \colhead{DDO 210 outer} }
\startdata 
$\bar\psi_{15-1}$ \hfill ($10^{-4}$ M$_\odot$yr$^{-1}$) & 0.44 & 0.29 \nl
$\bar\psi_{1-0}$ \hfill ($10^{-4}$ M$_\odot$yr$^{-1}$) & 1.14 & 0.09 \nl
$\bar\psi_{15-1}/A$ \hfill ($10^{-10}$ M$_\odot$yr$^{-1}$pc$^{-2}$) & 2.66 & 0.47 \nl
$\bar\psi_{1-0}/A$ \hfill ($10^{-10}$ M$_\odot$yr$^{-1}$pc$^{-2}$) & 6.87 & 0.15 \nl
\enddata
\end{deluxetable}


\end{document}